\documentclass[
5p
]{elsarticle}

\usepackage{lineno}
\usepackage[dvipdfmx]{hyperref}
\modulolinenumbers[5]

\journal{Journal of \LaTeX\ Templates}

\usepackage{amsmath,amssymb}
\usepackage{graphicx}
\usepackage{dcolumn}
\usepackage{bm}
\usepackage{color}
\biboptions{sort&compress}
\usepackage[normalem]{ulem}  
\def\br#1{\left( #1 \right)}
\def\bra#1{\left\{ #1 \right\}}
\def\brac#1{\left[ #1 \right]}

\def\non{\nonumber}

\begin{document}

\begin{frontmatter}

\title{
Lee-Yang zeros in lattice QCD for searching phase transition points 
}

\author[my1ad]{M.~Wakayama}
\address[my1ad]{Research Center for Nuclear Physics (RCNP), Osaka University, Ibaraki, Osaka 567-0047, Japan}
\author[my2ad,my3ad,my4ad]{V.~G.~Bornyakov}
\address[my2ad]{School of Biomedicine, Far Eastern Federal University, 690950 Vladivostok, Russia}
\address[my3ad]{Institute for High Energy Physics NRC Kurchatov Institute, 142281 Protvino, Russia}
\address[my4ad]{Institute of Theoretical and Experimental Physics NRC Kurchatov Institute, 117218 Moscow, Russia}
\author[my2ad,my4ad,my5ad]{D.~L.~Boyda}
\address[my5ad]{School of Natural Sciences, Far Eastern Federal University, 690950 Vladivostok, Russia}
\author[my2ad]{V.~A.~Goy}
\author[my2ad,my1ad,my6ad]{H.~Iida}
\address[my6ad]{Research and Education Center for Natural Sciences, Keio University, Hiyoshi 4-1-1, Yokohama, Kanagawa 223-8521, Japan}
\author[my2ad,my4ad]{A.~V.~Molochkov}
\author[my2ad,my7ad,my1ad]{A.~Nakamura}
\address[my7ad]{Theoretical Research Division, Nishina Center, RIKEN, Wako 351-0198, Japan}
\author[my2ad,my4ad,my8ad]{V.~I.~Zakharov}%
\address[my8ad]{Moscow Institute of Physics and Technology, Dolgoprudny, Moscow Region, 141700, Russia}

\begin{abstract}
We report Lee-Yang zeros behavior at finite temperature and density. 
The quark number densities, $\langle n \rangle$, are calculated at the pure imaginary chemical potential $i\mu_{qI}$, where no sign problem occurs. 
Then, the canonical partition functions, $Z_C(n,T,V)$, 
up to some maximal values of $n$ 
are estimated through fitting theoretically motivated functions to 
$\langle n \rangle$, 
which are used to compute the Lee-Yang zeros. 
We study the temperature dependence of the distributions of the Lee-Yang zeros 
around the pseudo-critical temperature region $T/T_c = 0.84$ - 1.35. 

In the distributions of the Lee-Yang zeros, we observe the 
Roberge-Weiss phase transition at $T/T_c \ge 1.20$. 
We discuss the dependence of the behaviors of Lee-Yang zeros 
on the maximal value of $n$, so that we can estimate a reliable infinite volume limit. 
\end{abstract}


\end{frontmatter}


\section{\label{intro}Introduction}

Revealing the phase structure at finite temperature and density in quantum chromodynamics (QCD) 
is one of the most important and interesting subjects in quark-hadron physics.
At high temperature and low density 
the quark-gluon plasma was created by heavy ion colliders at 
RHIC (BNL)~\cite{Adams:2005dq} and LHC (CERN)~\cite{Aamodt:2008zz}. 
High density region of the QCD phase diagram will be explored
by FAIR (GSI), NICA (JINR), and J-PARC (KEK/JAEA) in the near future. 
These experiments will provide us valuable information for understanding 
the early universe and the interior of neutron stars. 

On the theoretical side, 
the lattice QCD is an ideal tool to conduct directly non-perturbative calculations in QCD. 
The lattice QCD, however, suffers from the sign problem at finite density: 
the fermion determinant $\det D(\mu_q)$ at finite quark chemical potential $\mu_q$ is complex in general, 
and consequently,
it is impossible to apply the conventional Monte Carlo method. 

The canonical approach~\cite{Hasenfratz:1991ax}, which we use in this paper, 
is a promising candidate for avoiding the sign problem. 
Since the fermion determinant 
at pure imaginary chemical potential $\mu_q=i\mu_{qI}$ $(\mu_{qI}\in \mathbb{R})$ is real, 
canonical partition functions $Z_C(n,T,V)$ can be evaluated with the conventional Monte Carlo method at finite $\mu_{qI}$. 
Note that the canonical partition functions $Z_C(n,T,V)$ depend on 
a net number of quarks and antiquarks, $n$, temperature $T$ and volume $V$, 
but do {\it not} depend on the chemical potential.
Although the difficulty associated with the complex fermion determinant remains as the highly oscillating integral, 
authors of Refs.~\cite{Morita:2012kt,Fukuda:2015mva,Nakamura:2015jra} 
pointed out that the integral can be carried out with a multi-precision arithmetic. 
Once the canonical partition functions are extracted, 
the grand canonical partition function $Z_{GC}(\mu_q,T,V)$ with an arbitrary chemical potential can be constructed 
via the fugacity expansion, 
\begin{eqnarray}
Z _{\rm GC} (\mu_{q},T,V) &=& \sum_{n=-\infty}^{\infty} Z_C(n,T,V) \xi^n_q \ , \label{fuga_exp}
\end{eqnarray}
where $\xi_q = e^{\mu_q/T}$ is the quark fugacity. 
The canonical approach has been used in several analyses to reveal the QCD phase diagram~
\cite{Fukuda:2015mva,Nakamura:2015jra,deForcrand:2006ec,Ejiri:2008xt,Li:2010qf,Li:2011ee,Danzer:2012vw,Gattringer:2014hra,
Boyda:2017lps,Goy:2016egl,Bornyakov:2016wld,Boyda:2017dyo}.  

In this paper, we employ Lee-Yang zeros (LYZs) to clarify the QCD phase
structure.
Analysis of Lee-Yang zeros (LYZs) is a general and powerful tool
to investigate the phase structure. 
The theorems of Yang and Lee ~\cite{Yang:1952be,Lee:1952ig} state that 
zeros of the grand canonical partition function in complex fugacity plane
contain a valuable information on the phase transitions of a system. 

The number of LYZs is finite in a finite system. 
In the infinite size limit of the system, the number becomes infinite and one-dimensional cuts  are made by the zeros,
and
the edges of the cuts, which are the accumulation points of the LYZs,
are going to the phase transition point in the fugacity.
By investigating the behavior of the edges with system size, 
we can discuss the properties of phase transition.
 
In the literatures, there are several
studies of phase structure at finite density in lattice QCD 
by using LYZs.  
The pioneering work is that by Barbour and Bell~\cite{Barbour:1991vs}, 
which were performed on lattice volumes, $2^4$ and $4^4$.
Then the strong coupling calculation at zero temperature using the Glasgow algorithm~\cite{Barbour:1997bh}, 
and one dimensional QCD simulations~\cite{Lombardo:1999cz} were reported. 
Fodor and Katz studies QCD phase diagram by using LYZs on larger lattice sizes. 
They determined the critical point by distinguishing the first order transition and crossover 
with the multi-parameter reweighting method on realistic quark masses, $N_t=4$, and $N_s=6, 8, 10,$ and 12 lattices, 
where $N_t$ ($N_s$) is the number of lattice sites in temporal (spatial) direction~\cite{Fodor:2004nz}.
There was, however, a discussion of the reliability of their results~\cite{Ejiri:2005ts}. 

In Ref.~\cite{Nagata:2014fra}, LYZs are discussed both in analytic and theoretical approaches; 
the authors found that 
the scaling of LYZs with the system size in lattice QCD around the Roberge-Weiss (RW) phase transition point~\cite{Roberge:1986mm} 
is consistent with that derived analytically in the high temperature limit. 
In this paper, We will discuss on the behavior of LYZs around RW phase transition point.  
Recently, 
LYZs were estimated by the random matrix model~\cite{Morita:2015tma}. 

As shown in Ref.~\cite{Nakamura:2013ska}, 
the distributions of LYZs can be calculated 
from the experimental data of the net-proton multiplicity distribution at RHIC~\cite{Aggarwal:2010wy,Luo:2012kja}. 
Therefore, studies of LYZs of lattice QCD will be useful to compare with ones from data of future experiments.

In this paper,  we show the first results of the calculation of LYZs by 
the canonical approach. 
We discuss the dependence of distributions of LYZs at temperatures $T=0.84$-$1.35$ 
on $N_t=4$ and $N_s=8,$ 16, and 20 lattices 
with quark mass $m_\pi/m_\rho=0.80$.

It is difficult to calculate the grand canonical partition function 
$Z_{\rm GC}$ directly. 
Therefore we calculated $Z_{\rm GC}$ by the ``integration method"~\cite{Boyda:2017lps,Goy:2016egl,Bornyakov:2016wld,Boyda:2017dyo},  
in which $Z_{\rm GC}$ is obtained by integrating a fit function 
of baryon number densities $n_B$ in the pure imaginary chemical
potential regions. 
See Sec.~\ref{cano}. 

The method has the essential difference in calculating $Z_n$ from previous works. 
Since the number density is fit by some analytic functions and and 
we construct $Z_{\rm GC}$ from that,  
$Z_n$ can be calculable for infinitely large $n$ in principle.  

There is, however,  a practical limit of the maximum number of $Z_n$, $N_{\rm max}$, which is smaller than 
the size of the system. 
Consequently, we have to see the scaling of LYZs not only with the system size but also with $N_{\rm max}$.

The main objective of the present study is to investigate the distributions of LYZs for various temperatures from canonical method, 
and to extract information of the phase transition in real chemical potential region from the method. 

The organization of the paper is as follows.
In Sec.~\ref{cano}, the canonical approach is briefly described. 
In Sec.~\ref{Z_LYZ}, algorithms used in calculations of LYZs are introduced. 
Here ``cut Baum-Kuchen algorithm" is explained, which is used for finding the roots of the grand canonical partition function. 
In Sec.~\ref{simu}, the numerical results are shown. 
We show the properties of distributions of LYZs for temperatures $T=0.84$-$1.35$, 
and their behavior is examined for each temperature. 
Section~\ref{summ} is devoted to the summary.

\section{\label{cano}Canonical approach}

The canonical partition functions in Eq.~\eqref{fuga_exp} can be written as 
a Fourier transformation of the grand canonical partition function 
of the pure imaginary chemical potential~\cite{Hasenfratz:1991ax}, 
\begin{eqnarray}
\!\!\!\!\!\!\!\!
Z_C(n,T,V)  &=& \int_0^{2\pi} \frac{d\theta_q}{2\pi} \, e^{-in\theta_q} Z_{\rm GC} (i \mu_{qI},T,V) \ , \ \ \ 
\label{ZC}
\end{eqnarray}
where $\theta_q = \frac{\mu_{qI}}{T}$. 
We can construct $Z_{\rm GC}$ at the pure imaginary $\mu_q$ as
\begin{eqnarray}
\!\!\!\!\!\!\!\!\!\!\!\!
Z_{\rm GC} (i \mu_{qI},T,V) &=& C \exp{\bra{-V \int^{\theta_q}_0 d\theta_q^{\prime} \, n_{qI}(\theta_q^{\prime})}}  \ , \ \ \ \ \ 
\label{GCint}
\end{eqnarray}
where $C$ is an integration constant and $n_{qI}$ is an imaginary number density, $n_{q}=in_{qI}$. 
The number density $n_q$ is defined as 
\begin{eqnarray}
\!\!\!\!\!\!\!\!\!\!\!\!
 \frac{n_q}{T^3} &=& \frac{1}{VT^2}\frac{\partial}{\partial \mu_{q}} \ln Z _{\rm GC} (\mu_{q},T,V)  \non \\
                        &=&  \frac{N_f}{VT^3}\frac{1}{ Z_{\rm GC}} \int  {\cal D} U \br{\det D(\mu_q)}^{N_f} e^{-S_G}  \non \\
                          &&  \ \ \ \ \ \ \ \ \ \ \ \ \ \ \ \ \ \ \ \ \ \ \ \    \times {\rm Tr} \brac{D^{-1}\frac{\partial D}{\partial (\mu_{q}/T)}} \ , \ \ \ \   \label{nq_real}
\end{eqnarray}
where $S_G$ is a gauge action, $D(\mu_q)$ is a Dirac operator and $N_f$ is 
the number of flavors. 
The conventional Monte Carlo method is applicable 
to a calculation of $n_q$ at the pure imaginary $\mu_q$ 
since the fermion determinant is real. 

It was observed that the pure imaginary number density $n_{qI}$ can be well approximated by 
an odd power polynomial of $\theta_q$, 
\begin{eqnarray}
 \frac{n_{qI}}{T^3}(\theta_q) &=& \sum_{k=1}^{N_{{\rm poly}}} a_{2k-1} \theta^{2k-1}_q \ ,
 \label{poly}
\end{eqnarray}
at high temperature and by a Fourier series, 
\begin{eqnarray}
 \frac{n_{qI}}{T^3}(\theta_q) &=& \sum_{k=1}^{N_{{\rm sin}}} f_{3k} \sin(3k\theta_q) \ , 
 \label{mulsin}
\end{eqnarray}
at low temperature~
\cite{DElia:2016jqh,Gunther:2016vcp,Takahashi:2014rta,Takaishi:2010kc,DElia:2009pdy}. 
In this paper, we employ Eqs.~\eqref{poly} and \eqref{mulsin} 
with maximal values $N_{{\rm poly}}$ and $N_{{\rm sin}}$.

Note that when we fit Eq.~\eqref{mulsin} to the number density, 
the integration in Eq.~\eqref{GCint} can be performed as 
\begin{eqnarray}
\!\!\!\!\!\!\!\!\!\!\!\!
 Z_{GC}(i\mu_{qI},T,V)
 &=&  C \prod_{k=1}^{N_{\rm sin}} \exp \bra{   \tilde{f}_{3k} \cos \br{3k\theta_q} } \, , \ \ \ \ 
 \label{app1}
\end{eqnarray}
where $\tilde{f}_{3k}=\frac{N_s^3}{N_t^3}\frac{f_{3k}}{3k}$ and $C$ is an integration constant. 
Complex Fourier series of the parts of $Z_{GC}$ can be written 
with the modified Bessel function of the first kind, $I_n(x)$: 
\begin{eqnarray}
\!\!\!\!\!\!\!\!\!\!\!\!
 e^{ \tilde{f}_{3k} \cos \br{3k\theta_q} } &=& \sum_{n_k=-\infty}^{\infty} Z(3n_k,T,V;\tilde{f}_{3k}) e^{3 i n_k \theta_q} \, ,  \ \ \ 
 \label{app2} \\
\!\!\!\!\!\!\!\!\!\!\!\!\!\!\!
 Z(3n_k,T,V;\tilde{f}_{3k}) \!\! &=& \!\! \frac{3}{2\pi} \int_{-\frac{\pi}{3}}^{\frac{\pi}{3}} d\theta_q \, 
                                                                      e^{ \tilde{f}_{3k} \cos \br{3k\theta_q} } e^{-3 i n_k \theta_q}  \non \\
 &=& I_{n_k}(\tilde{f}_{3k}) \, .
 \label{app4}
\end{eqnarray}
Therefore, the grand canonical partition function follows the equation 
\begin{eqnarray}
\!\!\!\!\!\!\!\!\!\!\!\!\!\!\!
 Z_{GC}(\mu_q,T,V) \!\!&=&\!\! C \prod_{k=1}^{N_{\rm sin}}\br{\sum_{n_k=-\infty}^{\infty} I_{n_k}(\tilde{f}_{3k}) \xi_q^{3 n_k} } \, , \!
 \label{app5} \\
\!\!\!\!\!\!\!\!\!\!\!\!\!\!\!
 Z_{C}(3n,T,V) \!\!&=&\!\! C \!\!\!\!\!\! \sum_{\substack{ n_1, \cdots, n_{ N_{\rm sin} }  \\    =-\infty} }^{\infty} \!\!
                                   \brac{\br{\prod_{k=1}^{N_{\rm sin}} I_{n_k}(\tilde{f}_{3k})} \!
                                   \delta_{n, \sum_{j=1}^{N_{\rm sin}} n_j}}  \! ,   \non \!\! \\
                        &&
 \label{app6}
\end{eqnarray}
where all $Z_C(n)$ for $\mathrm{mod}(n,3)\neq 0$ are zero. 
In the case of $N_{\rm sin}=1$, it corresponds to the Skellam model~\cite{Skellam}: 
\begin{eqnarray}
 Z_{GC}(\mu_q,T,V) &=& C \sum_{n=-\infty}^{\infty} I_{n}(\tilde{f}_{3}) \xi_q^{3 n} \ , \label{app7} 
\end{eqnarray}
that is often used in analyzing energy heavy ion collision experiments.

\begin{table*}
\caption{
Simulation parameters for $m_{\pi}/m_{\rho} =  0.80$~\cite{Ejiri:2009hq} and  
results for fitting coefficients $f_{3k}$ or $a_{2k-1}$ from the data of $n_{qI}/T^3$ for each temperature.}
\begin{tabular}{ccc|cccccccc}\hline \hline
$\beta$ &$\kappa$ & $T/T_c$ &
$f_3$ & $f_6$ & $f_9$ & $f_{12}\times10$  & $f_{15}\times10^{2}$  & $f_{18}\times10^{2}$ & $f_{21}\times10^{2}$  & $N_{\rm dof}$\\\hline 
1.70 &0.142871 &0.84(4) & $0.084582$ & --- & ---  & --- & --- & --- & --- & 20 \\
1.80 &0.141139 &0.93(5) & $0.26139\ \, $ & --- & ---  & --- & --- & --- & --- & 40  \\
1.85 &0.140070 &0.99(5) & $0.73219\ \, $ & $-0.014036$ & $0.0034122$ & ---  & --- & --- & --- & 21  \\
1.90 &0.138817 &1.08(5) & $1.9801\ \ \, \,  $ & $-0.56241\ \, $ & $0.23489\ \ \, \, $ & $-1.0881$ & $5.1540$ & $-1.9676$ & $1.5413$  & 20 \\ \hline
$\beta$ & $\kappa$ & $T/T_c$ &
$a_1$ & $a_3$ & $a_5$ & $a_{7}$  & $a_{9}\times10^{2}$  &  &    & $N_{\rm dof}$\\ \hline
1.95 &0.137716 &1.20(6) & $4.4112$ & $-1.1175\ \: $ & 0.0000  & 0.0000 & $-8.1172$ &  &   & 19 \\
2.00 &0.136931 &1.35(7) & $4.6683$ & $-0.98663$ & --- & --- & --- &  &  & 26 \\\hline \hline

\end{tabular}
\label{table_coef}
\end{table*}

\begin{table}
\caption{
$T/T_c$ and $N_{\rm max}$ in our calculations.
The asterisks mean that the simulations include the negative $Z_{C}$.}
\begin{tabular}{cc|cc|cc} \hline \hline
$T/T_c$ &$N_{\rm max}$ & $T/T_c$ &$N_{\rm max}$ & $T/T_c$ &$N_{\rm max}$   \\\hline 
0.84&\ 50&  1.08&\ 25& 1.20&\ 25 \ \\
       &100&         &\ 50&        &\ 50 \ \\
       &150&         &\ 75&        &\ 75 \ \\
       &200&         &100&        &\ 95 \ \\
\cline{1-2}
0.93&\ 50&         &125&        &100$^{\ast}$ \\ 
\cline{5-6}
       &100&         &150& 1.35&\ 25 \ \\
       &150&         &200&        &\ 50 \ \\
       &200&         &250&        &\ 75 \ \\
\cline{1-2}
0.99&\ 50&         &300&        &100 \, \\
       &100&         &      &        &125$^{\ast}$ \\
       &150&         &      &        &150$^{\ast}$ \\
       &200&         &      &        &200$^{\ast}$ \\ \hline \hline
\end{tabular}
\label{table_Nmax}
\end{table}

\section{\label{Z_LYZ}Partition functions and Lee-Yang zeros}

The grand canonical partition function can be rewritten as 
\begin{eqnarray}
\!\!\!\!\!\!\!
Z _{\rm GC} (\mu_{B}/3,T,V) &=&  \sum_{n=-N_{\rm max}}^{N_{\rm max}}  Z_{C}(3n,T,V) \xi^n _B \ ,
\label{Zbaryon}
\end{eqnarray}
where we took into account the property of the canonical partition function 
$Z_C(n,T,V)=0$ for $\mathrm{mod}(n,3)\neq 0$ 
because of the Roberge-Weiss symmetry on the number density.
Here $\mu_B = 3\mu_q$ is the baryon chemical potential and $\xi_B=\xi_q^3$ is the baryon fugacity. 
We truncate the fugacity expansion at $|n|=N_{\rm max}$. 
On an $N_s^3\times N_t$ lattice at finite temperature, 
where $N_s$ ($N_t$) is the number of lattice sites in spatial (temporal) direction, 
the exact grand canonical partition function is obtained for $N_{\rm max} = 2N_s^3N_f$. 

Lee-Yang zeros, $\alpha_n$, are given as roots of the equation, 
\begin{eqnarray}
\!\!\!\!\!\!\!\!\!\!\!\!\!\!\!
  f(\xi_B=\alpha_n) &\equiv& \xi_B^{N_{\rm max}} \!\!\!\! \sum_{n=-N_{\rm max}}^{N_{\rm max}} \!\! Z_{C}(3n,T,V) \xi^n _B  =  0 \ , \ \ \ 
\label{LYeq1}
\end{eqnarray}
in the complex $\xi_B$ plane. 
We must solve a polynomial equation of high degree of $2N_{\rm max}$.
This is a famous ill-posed problem.
In Ref.~\cite{Nakamura:2013ska}, a new algorithm was proposed to overcome the difficulty:
the cut Baum-Kuchen (cBK) algorithm with the multi-precision arithmetic. 
The multi-precision arithmetic is done with the FMLIB package~\cite{FMLIB} with 100-300 significant digits.

\begin{figure}
\begin{center}
\includegraphics[width=0.46\linewidth]{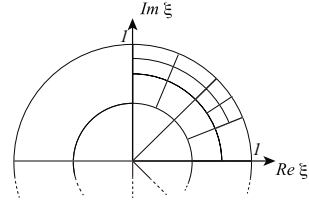}
\vspace{-8mm}
\caption{
Schematic diagram to show paths in the cut Baum-Kuchen algorithm, taken from Ref.~\cite{Nakamura:2013ska}.
We perform the Cauchy's integration, Eq.~\eqref{cauchy},  
along a closed path, which gives the number of LYZs inside of the path.
If the outcome is not zero, we divide the path into four pieces and
continue the integration on each path.
}
\label{cBK}
\end{center}
\end{figure}

The cBK algorithm is as follows. 
Since the $\alpha_n$ are roots of $f(\xi_B)$,
\begin{eqnarray}
f(\xi_B) &\propto &  \prod_{n=1}^{2N_{\rm max}} (\xi_B - \alpha_n).
\end{eqnarray}
It is easy to see that $f(\xi_B)$ satisfies
\begin{eqnarray}
 \frac{f^{\prime}(\xi_B)}{f(\xi_B)}&=& \sum_{n=1}^{2 N_{\rm max}} \frac{1}{\xi_B - \alpha_n} . 
 \label{ff}
\end{eqnarray}
Then, we can extract the number of LYZs 
inside a closed-integral path $C_0$
by Cauchy's integral,
\begin{eqnarray}
 \frac{1}{2\pi i}\oint_{C_0} \frac{f^{\prime}(\xi_B)}{f(\xi_B)}\,  d\xi_B \ . 
 \label{cauchy}
\end{eqnarray}
$Z_C(n,T,V)$ satisfy 
\begin{eqnarray}
 Z_C(+n,T,V) &=& Z_C(-n,T,V) \ ,
\label{Zc+-}
\end{eqnarray}
from the charge-parity invariance. 
If $\alpha_n$ is a LYZ, then $\alpha_n^{\ast}$ and $\alpha_n^{-1}$ are also LYZs
because the $Z_C(n,T,V)$ are real and satisfy Eq.~\eqref{Zc+-}. 
Thanks to these properties, it is enough to search only inside the upper half of the unit circle 
in the complex $\xi_B$ plane. 
The first closed-integral path $C_0$ 
is along the boundary of the plane shape defined by relations 
$0 \le {\rm arg} \br{\xi_B} \le \pi $ and $0 \le |\xi_B| \le 1$. 
If there are LYZs inside a contour, 
we divide respective annulus sector  into four parts bisecting  both radial and angular ranges of polar coordinates 
and perform the Cauchy integrals for each sector. 
In the cBK algorithm we can locate LYZs by conducting this procedure recursively.
See Fig.~\ref{cBK}.

\section{\label{simu} Lattice QCD simulations}
\subsection{\label{setup} Lattice setup}

We generate the gauge field configurations in full QCD using the hybrid Monte Carlo method~\cite{Duane:1987de} 
with the Iwasaki gauge field action~\cite{Iwasaki:1985we} 
and the $N_f=2$ clover improved Wilson fermion action~\cite{Sheikholeslami:1985ij}.
Simulations are carried out on a $N_s^3\times N_t = 16^3\times4$ lattice 
at temperatures $T/T_c=0.84,$ 0.93, 0.99, 1.08, 1.20, and 1.35 with $m_{\pi}/m_{\rho}=0.80$. 
We also conduct simulations on $N_s^3=8^3$ (for $T/T_c=0.93$ and 1.35) and $N_s^3=20^3$ (for $T/T_c=1.35$) lattices
to check the volume dependence. 
We use the same parameters (couplings $\beta$ and hopping parameters $\kappa$) 
as those in Ref.~\cite{Ejiri:2009hq}. 
The pseudo-critical temperature $T_c$ was determined from the peak of the Polyakov-loop susceptibility~\cite{Ejiri:2009hq}. 
All parameters on a $N_s^3=16^3$ lattice are listed in Table~\ref{table_coef}. 
The clover coefficient $c_{SW}$ in the clover improved Wilson fermion action 
is given by $c_{SW} = (1-0.8412\beta^{-1})^{-3/4}$~\cite{Iwasaki:1985we}. 
We produce 4000 (2000) gauge field configurations taking every 10th trajectory 
for $T/T_c=0.84,$ 0.93, 0.99, and 1.08 ($T/T_c=1.20$ and 1.35). 
The first 200 configurations are discarded for thermalization. 

\subsection{\label{LYZxi} Lee-Yang zeros in the complex $\xi_B$ plane}

We calculate the number density $n_{qI}/T^3$ at 19 to 40 values of $\mu_{qI}$ depending on temperature. 
We fit the $n_{qI}/T^3$ with Eq.~\eqref{mulsin}
with $N_{\rm sin}=1,$ 1, 3, and 7 for $T/T_c=0.84,$ 0.93, 0.99, and 1.08, 
and with polynomials Eq.~\eqref{poly} 
with $N_{\rm poly}=5$ and 2 for $T/T_c=1.20$ and 1.35, respectively. 
Coefficients, $f_k$ and $a_k$, obtained by the fitting are shown in Table~\ref{table_coef}. 

As shown in Table~\ref{table_Nmax}, 
we calculate LYZs with the cBK algorithm 
for few values of $N_{\rm max}$ for each temperature value.
After the cBK recursions are carried out with eight or more times, 
we can obtain  
magnitude $r_B=|\xi_B|$ and argument $\theta_B = {\rm arg} \br{\xi_B}$ of LYZs 
at centers of obtained annulus sectors. 
Due to the finite size of the annulus sectors, 
the LYZ coordinates have systematic errors: 
$\delta r_B< 2.0\times10^{-3} \sim 1/2^9$ and 
$\delta \theta_B<6.2\times10^{-3} \sim \pi/2^9$. 

\begin{figure}
\begin{center}
\includegraphics[scale=0.46]{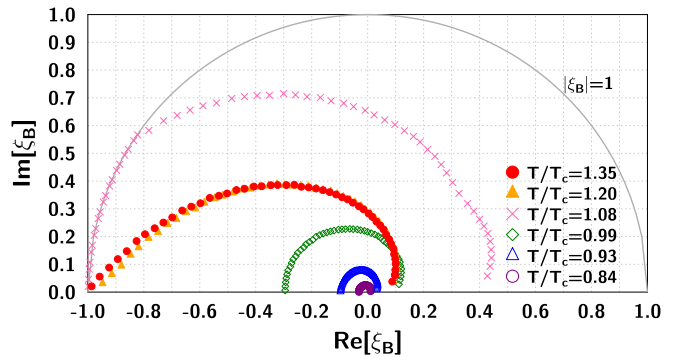}
\caption{ (color online). 
The temperature dependence of LYZs in the complex $\xi_B$ plane. 
$N_{\rm max}$ is 100 for all temperature except for $T/T_c=1.20$. 
For $T/T_c=1.20$ $N_{\rm max}$ is 95. 
The solid line stands for the unit circle. 
}
\label{Tdep}
\end{center}
\end{figure}

\subsubsection{\label{xi_temp} Temperature dependence}

In Fig.~\ref{Tdep}, we present the temperature dependence of LYZs. 
We show in Figs.~\ref{Tdep}, \ref{b200}, \ref{b180}, and \ref{b190} 
only part of the LYZs. 
The rest of them can be restored using complex conjugation or inversion; 
See the previous section.
For all temperatures except $T/T_c=1.20$, LYZs are calculated with $N_{\rm max} =100$. 
The calculation of LYZs for $T/T_c=1.20$ is terminated at $N_{\rm max} =95$ 
because for $T/T_c=1.20$ negative values of $Z_{C}(3n,T,V)$ appear for $3n\ge 294$, 
see also Ref.~\cite{Boyda:2017dyo}. 
Note that all LYZs have the systematic errors within symbols 
although the errors are not displayed in Fig.~\ref{Tdep}.
The same for the following figures. 

Distributions of LYZs above $T_c$ and below $T_c$ are quite different. 
Below $T_c$ the distributions of the LYZs are nearly circles. 
And as the temperature becomes lower, the radius of the distribution becomes smaller for the fixed $N_{\rm max}$. 

On the other hand, above $T_c$ the distributions are reaching to the value $\xi_B=-1$. 
In the complex $\mu_q/T$ plane, this point corresponds to 
\begin{eqnarray}
\xi_B = -1  \ &\Leftrightarrow& \ \mu_{q}/T = \frac{(2k+1)\pi i}{3} \ , 
\end{eqnarray}
where $k$ is an integer. 
Thus, 
$\xi_B=-1$ represents 
the RW phase transition. 
At $T/T_c=1.35,$ 1.20, and 1.08, there are LYZs that are very close to $\xi_B=-1$.
We do not see, however, a LYZ exactly at $\xi_B=-1$. 
This is probably due to finite $V$ and finite $N_{\rm max}$ effects.

\subsubsection{\label{xi_Nmax}V and $N_{\rm max}$ dependences}

Phase transitions of QCD shall be seen in large-$V$ and large-$N_{\rm max}$ limits. 
Consequently, it is important whether a LYZ appears on the real positive axis in the complex $\xi_B$ plane 
as $V$ and $N_{\rm max}$ go to large values. 

\begin{figure}
\begin{center}
\includegraphics[scale=0.46]{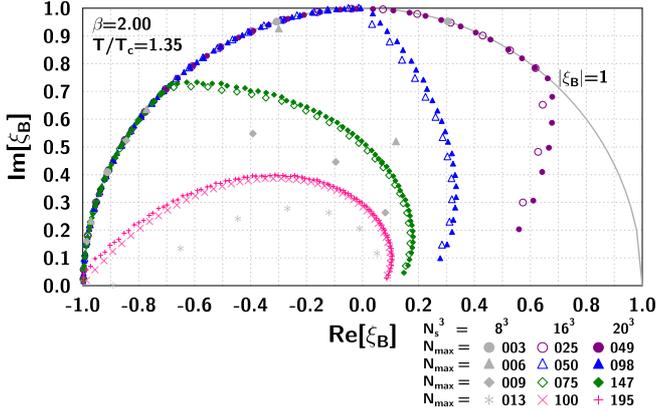}
\caption{ (color online). 
The $V$ and $N_{\rm max}$ dependences of LYZs at $T/T_c=1.35$ in the complex $\xi_B$ plane. 
The solid line represents the unit circle. 
}
\label{b200}
\end{center}
\end{figure}

\begin{figure}
\begin{center}
\includegraphics[scale=0.46]{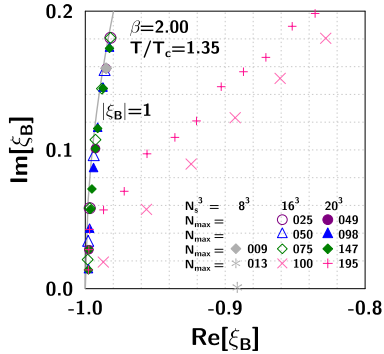}
\caption{ (color online). 
LYZs near $\xi_B=-1$ in Fig.~\protect\ref{b200}. 
The solid line represents the unit circle. 
}
\label{b200_zoom}
\end{center}
\end{figure}

\begin{figure}
\begin{center}
\includegraphics[scale=0.598]{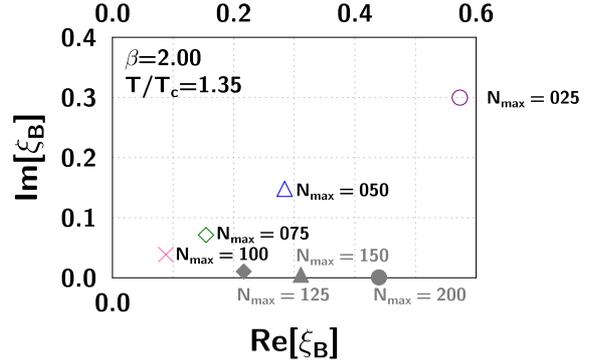}
\caption{ (color online). 
The $N_{\rm max}$ dependence of the right edges of LYZs at $T/T_c=1.35$ in the complex $\xi_B$ plane. 
The results above $N_{\rm max} = 125$ are polluted by negative values of $Z_{C}(3n,T,V)$. 
}
\label{edge}
\end{center}
\end{figure}

\begin{figure}
\begin{center}
\includegraphics[scale=0.46]{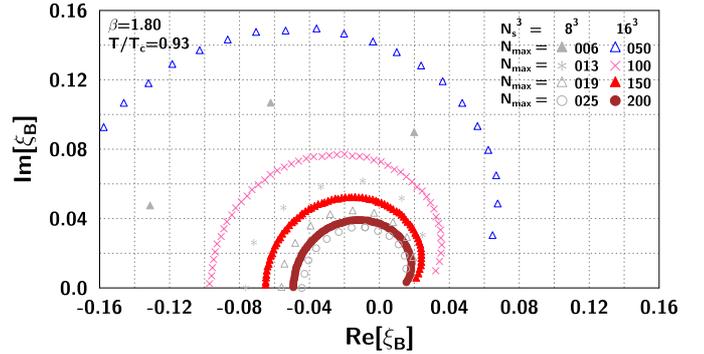}
\caption{ (color online). 
The $V$ and $N_{\rm max}$ dependences of LYZs at $T/T_c=0.93$ in the complex $\xi_B$ plane. 
}
\label{b180}
\end{center}
\end{figure}

\begin{figure}
\begin{center}
\includegraphics[scale=0.46]{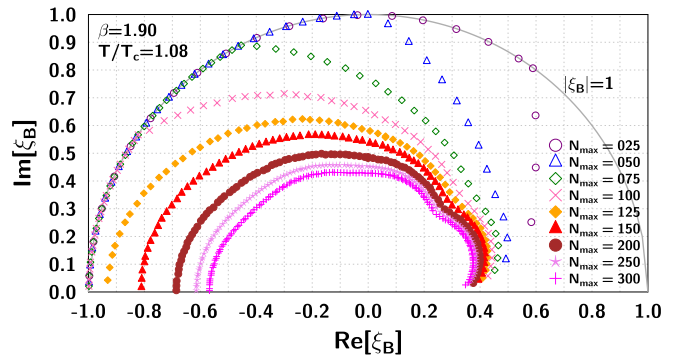}
\caption{ (color online). 
The $N_{\rm max}$ dependence of LYZs at $T/T_c=1.08$ in the complex $\xi_B$ plane. 
The solid line represents the unit circle. 
}
\label{b190}
\end{center}
\end{figure}

In Fig.~\ref{b200}, 
we show the $V$ and $N_{\rm max}$ dependences of LYZs at $T/T_c=1.35$ 
in the complex $\xi_B$ plane. 
Following Ref.~\cite{Nakamura:2013ska}, 
we choose $N_{\rm max}$ so that the value of $N_{\rm max}/V$ is approximately same when $V$ changes. 
It means that we fix number of baryons that can exist in the system per volume. 
Then, we find that LYZs fall onto the almost same curve if $V$ is sufficiently large as $16^3$ or $20^3$. 

There are some LYZs on the unit circle at small-$N_{\rm max}$. 
However, they go away from the unit circle when $N_{\rm max}$ becomes larger: 
thus these LYZs are an artifact due to the small $N_{\rm max}$.
As can be seen in Fig.~\ref{edge},
when we increase $N_{\rm max}$, the right edge of LYZs goes to zero, 
where the right edge is determined by a position of the LYZ 
for ${\rm Re} \brac{\xi_B} > 0$ and $\min \br{{\rm Im} \brac{\xi_B}}$.
Therefore, there is no phase transition for real $\mu_B/T$ at $T/T_c=1.35$. 

On the contrary, an important observation is that 
the LYZ nearest to $\xi_B=-1$ approaches this point 
as $V$ increases 
and the LYZ is stable with respect to changing of $N_{\rm max}$; 
See Fig.~\ref{b200_zoom}. 
This tells us that there is a RW phase transition at $T/T_c=1.35$. 

We calculated LYZs also for $N_{\rm max} \ge 125$.
But the results are polluted by negative values of $Z_{C}(3n,T,V)$ 
appearing from $3n=312$. 
As $N_{\rm max}$ increases beyond $N_{\rm max} = 125$, the right edge of LYZs suddenly turns away from zero 
in spite of approaching zero until $N_{\rm max} = 100$, see Fig.~\ref{edge}. 
The distributions of the LYZs are sensitive to the inclusion of the negative $Z_C$. 

Next, we investigate the $N_{\rm max}$ dependence of LYZs in the confinement phase. 
In Fig.~\ref{b180}, the result at $T/T_c=0.93$ is presented. 
Simulations are carried out up to $N_s=16$ 
because we find that the $N_s=16$ lattice is sufficiently large in Fig.~\ref{b200}. 
In the following figures, we only show results for $N_s=16$ and focus the discussion on the $N_{\rm max}$ dependence. 
In Fig.~\ref{b180}, 
as $N_{\rm max}$ increases, a right edge of LYZs approaches to the real positive axis. 
We extrapolate the right edges of LYZs to the real axis by linear or quadratic functions 
and a phase transition point can be roughly estimated as $\mu_B/T\sim 5$-6 at $T/T_c=0.93$. 
At $T/T_c=0.99$, we also estimate a phase transition point  $\mu_B/T\sim 3$-3.5 with the same determination. 
We should point out that the determination of a phase transition strongly depends on a fitting function. 
Thus, we do not mention an estimation at $T/T_c=0.84$ 
because right edges of LYZs at $T/T_c=0.84$ are too close to zero: $\min(|\xi_B|)=0.0051$. 

The $N_{\rm max}$ dependence of LYZs at $T/T_c=1.08$ in the complex $\xi_B$ plane 
is shown in Fig.~\ref{b190}. 
At $T/T_c=1.08$, the $Z_{C}(3n,T,V)$ for all $n$ are positive. 
Because 
LYZs near $\xi_B=-1$ do not exist 
for $N_{\rm max}$ larger than or equals to 125, 
there is no RW phase transition at $T/T_c=1.08$. 
The right edge of LYZs approaches the real positive axis as $N_{\rm max}$ increases. 
However, the right edge is still far from zero even for $N_{\rm max}=300$; 
This suggests that there might be a stable point around 
$\mu_B/T \sim$ (1.03,0.08).

\subsection{\label{LYZmuT} Lee-Yang zeros in the complex $\mu_B/T$ plane}

\begin{figure}
\begin{center}
\includegraphics[scale=0.48]{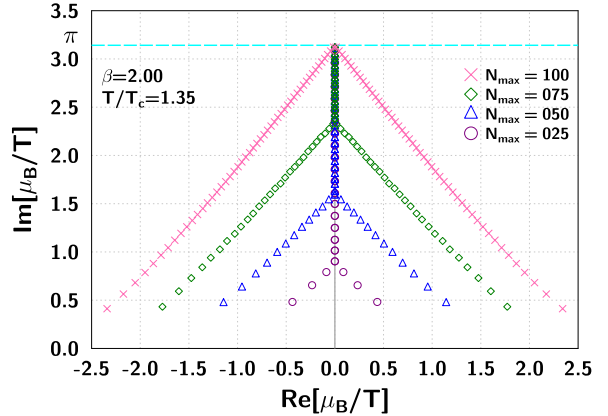}
\caption{ (color online). 
The $N_{\rm max}$ dependence of LYZs at $T/T_c=1.35$ in the complex $\mu_B/T$ plane. 
The solid line corresponds to the unit circle in Fig.~\protect\ref{b200}.
The dashed line represents the RW symmetry line.
}
\label{b200muT}
\end{center}
\end{figure}

Let us study the distributions of LYZs in the complex $\mu_B/T$ plane. 
This includes the same information as in the complex fugacity plane, but allows us
to see LYZ structure from different perspective. 
Fig.~\ref{b200muT} shows the results of the LYZs in the complex $\mu_B/T$ plane at $T/T_c=1.35$. 
The LYZs shown in Fig.~\ref{b200} correspond to ones in the second quadrant in Fig.~\ref{b200muT}. 
Since LYZs have the properties discussed above, 
LYZs also exist in the first quadrant. 
We calculated the LYZs also at $N_{\rm max} = 125,$ 150, and 200 
but they suffered from the negative $Z_C$.

We find that the LYZs far from the imaginary axis 
fall onto approximately a straight line 
for each $N_{\rm max}$. 
In other words, the LYZs in the upper unit circle in Fig.~\ref{b200} locate on a curved line, 
\begin{eqnarray}
\xi_B &=& e^{-t} e^{i(pt+\pi)} \ ,
 \label{curve}
\end{eqnarray}
with a parameter $t$ and two constants $p$ and $q$.

\section{\label{summ}Summary}

We studied LYZs, i.e., zeros of the grand canonical partition function, 
which we calculated from the number density using the integration method. 
The number density at the pure imaginary chemical potential were evaluated 
numerically in two-flavor full QCD. 
We used the Iwasaki gauge action and the clover improved Wilson fermion action. 
Simulations were carried out on $N_t=4$ and $N_s=8,$ 16, and 20 lattices at $m_{\pi}/m_{\rho}=0.80$ 
for temperatures in the range $T/T_c=0.84$ - 1.35. 

We obtained the canonical partition functions, $Z_C(n,T,V)$, 
after fitting the number density with Fourier series with $N_{\rm sin}=1,$ 1, 3, 
and 7 for $T/T_c=0.84,$ 0.93, 0.99, and 1.08, 
and with polynomials with $N_{\rm poly}=5$ and 2 for $T/T_c=1.20$ and 1.35. 

We obtained LYZs with the cut Baum-Kuchen algorithm for each temperature 
at few $N_{\rm max}$ values. 
The Roberge-Weiss phase transition clearly appears in the distribution of LYZs 
at $T/T_c=1.35$ and 1.20. 
Additionally, in the complex $\mu_B/T$ plane, 
the LYZs far from the imaginary axis 
fall onto approximately straight line 
for each $N_{\rm max}$ at $T/T_c=1.35$ and 1.20. 

We estimated phase transitions in the confinement phase as 
$\mu_B/T\sim 3$-3.5 at $T/T_c=0.99$ and 
$\mu_B/T\sim 5$-6 at $T/T_c=0.93$. 
More precise values at these temperature and an estimation at $T/T_c=0.84$ 
need calculations with larger $N_{\rm max}$. 

Moreover, in the deconfinement phase near $T_c$ ($T/T_c=1.08$), 
it is likely that there is a stable point around 
$\mu_B/T \sim$ (1.03,0.08) 
with respect to changing of $N_{\rm max}$. 
To confirm this, 
we need study more extensively the relation not only between the stable point and $N_{\rm max}$, 
but also between the point and $N_{\rm sin}$ below $T/T_c=1.08$. 

Above $T/T_c=1.20$, 
the negative $Z_C(n,T,V)$ appear for large-$n$, which
should not be used in the search of phase transitions from LYZs. 
Therefore, our searches of the $N_{\rm max}$ dependence are limited in some $N_{\rm max}$.  
To avoid the negative $Z_C$, we need more statistics 
and extract fitting coefficients $a_{2k-1}$ of a polynomial more accurate.

\section*{acknowledgments}
One of the authors (M.W.) wishes to thank M.~Sekiguchi for discussions and encouragement. 
This work was completed due to support by the Russian Science Foundation Grant under Contract No. 15-12-20008. 
Work done by A. Nakamura on the theoretical formulation of the $Z_n$ 
was supported by JSPS KAKENHI 
Grant Numbers 26610072 and 15H03663. 
Computer simulations were performed on 
the GPU cluster Vostok-1 
of Far Eastern Federal University. 
This research used computational resources of 
Cybermedia Center of Osaka University through the HPCI System (ID:hp170197). 
This work was supported by 
``Joint Usage/Research Center for Interdisciplinary Large-scale Information Infrastructures" in Japan (ID: EX18705).

\section*{References}

\bibliography{mybibfile}

\end{document}